# Design and Tests of 500kW RF Windows for the ITER LHCD System

J. Hillairet[1], J.Kim[2], N.Faure[3], J.Achard[1], Y.S.Bae[2], J.M.Bernard[1], L.Delpech[1], M.Goniche[1], S.Larroque[1], R.Magne[1], L.Marfisi[1], S.Park[2], S.Poli[1], N.Dechambre[3], K.Vulliez[4]

1. CEA, IRFM, F-13108 Saint Paul-lez-Durance, France.
2. National Fusion Research Institute, Daejeon, Korea.
3. PMB/ALCEN - Site de Peynier - Route des Michels, 13790 Peynier, France.
4. Laboratoire d'étanchéité, CEA DEN/DTEC/SDTC, 2 rue James Watt 26700 Pierrelatte, France.

## Abstract

In the frame of a R&D effort conducted by CEA toward the design and the qualification of a 5 GHz LHCD system for the ITER tokamak, two 5 GHz 500 kW/5 s windows have been designed, manufactured and tested at high power in collaboration with the National Fusion Research Institute (NFRI). The window design rely on a symmetrical pill-box concept with a cylindrical beryllium oxide ceramic brazed on an actively water cooled copper skirt. The ceramic RF properties have been measured on a test sample to get realistic values for guiding the design. Low power measurements of the manufactured windows show return losses below -32 dB and insertion losses between -0.01 dB and -0.05 dB, with an optimum frequency shifted toward lower frequencies. High power tests conducted at NFRI show unexpected total power loss for both windows. The ceramic temperature during RF pulses has been found to reach unexpected high temperature, preventing these windows to be used under CW conditions. A post-mortem RF analysis of samples taken from one window shows that the dielectric properties of the ceramic were not the ones measured on the manufacturer sample, which partly explain the differences with the reference modelling.

## 1. Introduction

The high current drive efficiency of lower hybrid waves makes Lower Hybrid Current Drive (LHCD) system a crucial actuator to sustain a fraction of the non-inductive current drive in tokamaks. Simulations of ITER scenarios show that LHCD can help saving the poloidal flux consumption during ramp-up phases and thus extends significantly the flat-top for the burning phase [1] . On the technological aspect, if the viability of the Passive Active Multijunction launcher concept has been validated in steady-state conditions at 3.7 GHz [2], its scalability to 5 GHz especially in terms of high power handling and Continuous Wave (CW) operations, remain to be demonstrated. Present LH launchers in Alcator C-Mod, EAST, KSTAR, work at frequencies of 4.6 GHz or 5 GHz. But none of them are actively cooled nor achieved total coupled Radio Frequency (RF) power higher than 2 MW CW. In this context and even if the LHCD system is not part of the construction baseline, the CEA/IRFM is conducting a R&D effort in order to validate the design and the performances of the critical RF devices of the 5 GHz ITER LHCD system.

A tokamak operates at pressures different from the transmission line pressure. RF windows intend to isolate these different pressures but allow propagation of the microwaves with low losses and limited RF power reflection. Vacuum vessel operates at ultra-low pressure, typically $10^{-4}$ to $10^{-6}$ Pa, while transmission lines can contain atmospheric pressure or pressurized gas to increase the breakdown voltage. Consequently, the window must be sufficiently strengthened to withstand differential pressures of the order of several tens of kPa. On the other hand, window must withstand mechanical stresses such as shock and vibration, and important temperature variations. In the case of ITER, the windows are part of the first confinement barrier.

However, the power handling capability of LH launchers is limited by multipactor [3] or non-linear coupling effects [3][4]. Moreover, as the dimensions of the tokamak access ports (or access port-plugs in the case of ITER) are limited, the number of transmission lines connected to the antenna is limited too. In order to fit into the available space, to simplify the assembly design and to improve the maintenance of the system, the number of transmission lines and thus the number of RF windows must be kept as low as possible. This leads to increase the power handling capability of each transmission line, and ultimately of each window.

In the current ITER LHCD design, 24 MW CW of RF power at 5 GHz are expected to be generated and transmitted to the plasma. In order to separate the vacuum vessel pressure from the cryostat waveguide pressure, forty eight 5 GHz 500 kW windows are to be assembled on the waveguides at the equatorial port flange. For nuclear safety reasons, forty eight additional windows could be located in the cryostat section, to separate the cryostat waveguide pressure from the exterior transmission line pressure and to monitor it. Because the failure of a window would lead to lock a transmission line, these windows have been identified as one of the main critical components for the ITER LHCD system since first ITER LHCD studies [5], and clearly require an important R&D effort [6] [7] [8] . In order to initiate this R&D effort, two 5 GHz 500 kW/5 s pill-box prototype windows have been developed and manufactured in 2012 by the PMB/ALCEN Company in close collaboration with the CEA/IRFM [9].

The section 2 of this paper reports the RF and the thermo-mechanical design of these 5 GHz BeO windows. Experimental RF measurements at low power are reported in section 3. High power results, obtained in collaboration with National Fusion Research Institute (NFRI), are reported in section 4 and analyzed in section 5.

## 2. RF and Thermo-Mechanical Design

The proposed ITER 5 GHz RF window is based on a pill-box window concept [10][11], i.e. a thin ceramic disc brazed in the middle of a short straight section of a circular waveguide axially connected on both sides to rectangular waveguides. According to the transmission line theory, the pill-box window has four discontinuities: rectangular waveguide to circular waveguide, vacuum to ceramic and ceramic to vacuum and circular waveguide to rectangular waveguide. Typical design rule of thumb of such device is circular section diameter about the same size of the diagonal of the rectangular waveguide. Without taking into account the ceramic, the circular section length is approximately half a guided wavelength of the circular $TE_{11}$ mode, in order to not generate spurious reflection into the rectangular sections. For a prescribed dielectric properties, the RF design consists in optimizing the geometrical dimensions of the window, mainly the diameter, the ceramic thickness and the vacuum circular waveguide length in order to minimize both RF reflection (matching) and RF power absorption in the ceramic (from trapped or ghost-mode heating [10] ). Once optimized, taking into account the ceramic, the matching is optimum only for a narrow band of frequency and is very sensitive to the device dimensions and the ceramic relative permittivity. As the klystrons used for fusion applications have narrow bandwidth, for instance 5 MHz at 3.7 GHz[12] [13] or 5 GHz klystrons [14] [15] , the RF optimization is performed for the central frequency only.

The heat losses in the ceramic, which have to be extracted by an active water cooling, depend on both the inside electric field topology and the ceramic dielectric loss (loss tangent). Undesirable modes due to parasitic resonances can be excited in the ceramic volume, raising the electric field and thus the heat dissipation. This aspect is uncorrelated from the return loss and one can even achieve low return loss but can have high heat losses in the ceramic (>2 kW) [20]. So both aspects have to be taken into account during the RF optimization process.

In the 2001 initial Lower Hybrid (LH) antenna design [16] [17] , 24 RF windows are set on the cryostat port. Each of them feeds 2 RF windows installed on the vacuum port. Since the antenna is designed for a power up to 20 MW, each RF window of the first and second group have to withstand respectively 833 kW CW and 417 kW CW. Two preliminary designs had been proposed in [16]. However, the mechanical analysis performed in [19] showed that stresses are higher than the static fatigue limit of 50 MPa. This particularly point invalidates the design for dynamic fatigue (cycles of repeated heating and cooling). Following this study, a new RF design has been developed. At the contrary of the model proposed in [16] [17] , a larger rectangular input section has been used (WR229 instead of WR187). It was found in [7] a range of dielectric thickness which combines low total dielectric losses (<650W) by minimizing the axial electric field in the ceramic and a low VSWR ($S_{11}$<-25dB). We did not take into account in this study the impact of non-negligible amount of reflected power (>5%), such as experimented on tokamak.

As highlighted in [21], the relative dielectric properties of the ceramic disk depend on the manufacturer, even when the same manufacturer and ceramic batch are used. At the frequency of 5 GHz, the dielectric losses (loss tangent) of aluminium oxide ($Al_2O_3$), aluminium nitride (AlN) or Beryllium Oxyde (BeO) are of the same order (tan δ ~ $10^{-4}$). The choice for BeO ceramic is motivated by its larger thermal conductivity (in the range of 330 to 370 W/m/K at room temperature) than the other ones (in the range of 30 W/m/K for $Al_2O_3$ and 70-180 W/m/K for AlN). BeO is used in high power 3.7 GHz and 5 GHz klystrons windows. In order to determine the final thickness of the windows, the BeO ceramic RF properties must be known with accuracy. A BeO sample has been requested from the ceramic manufacturer (American Beryllia) with the guarantee that the final ceramics will be made from the same lot. The sample has been characterized within a RF resonant cavity setup in collaboration with the XLIM laboratory [25]. The measurement of the BeO sample has been made at two surrounding frequencies and the results are given in **Table 1**. The relative permittivity and the loss tangent at 5 GHz have been calculated from a linear interpolation of the measured values, to be $ε_r$=6.36±0.159 and tan δ = 4.30e-4±3.82e-5 respectively.

| Measurement Frequency | 4.7 GHz | 5.31 GHz |
|---|---|---|
| Relative permittivity ($ε_r$) | 6.37 ± 0.153 | 6.35 ± 0.165 |
| Loss tangent (tan δ) | 3.68e-4 ± 3.73e-5 | 4.93e-4 ± 3.91e-5 |

**Table 1. Measurements results of the BeO sample of American Beryllia from XLIM laboratory.**

From these measurements, the final dimensions have been calculated by the PMB/ALCEN Company in collaboration with CEA. In the proposed design, a rectangular waveguide of section 58.17x29.08 mm (WR229 standard) is used to forward the RF power into the circular section. The ceramic disk diameter and the thickness are 85.89 mm and 8.3 mm respectively. Other dimensions are reported in **FIGURE 1**. In order to increase the RF performance of the window, in particular the return loss, two matching inductive rods have been added on each side of the window. These rods also make possible an eventual post-manufacturing tuning, by slightly deforming them after the final assembly. This technique has also been used in the TED 3.7 GHz windows used on the Tore Supra tokamak, but with a larger rod diameter due to the reduced frequency.

The power loss in the ceramic volume and the electric field topology are illustrated in Figure 2. With this design, the calculated dissipated power in ceramic is around 550 W for 500 kW input on matched load. Taking into account a conservative copper conductivity of 4.4 MS/m, the total resistive losses are estimated to 720 W. Combined, the expected total RF losses for the window to evacuate is thus of the order of 1.27 kW.

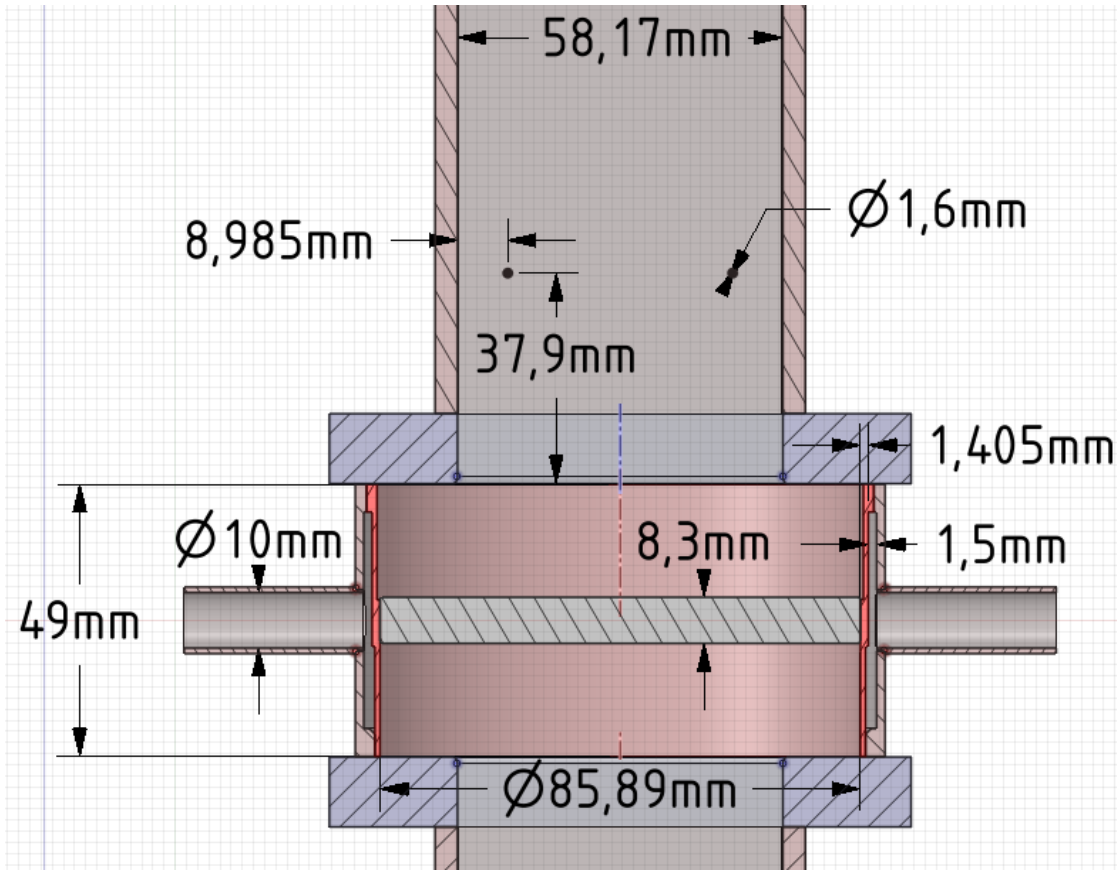

**FIGURE 1**. Main dimensions of the final 5 GHz RF window prototype.

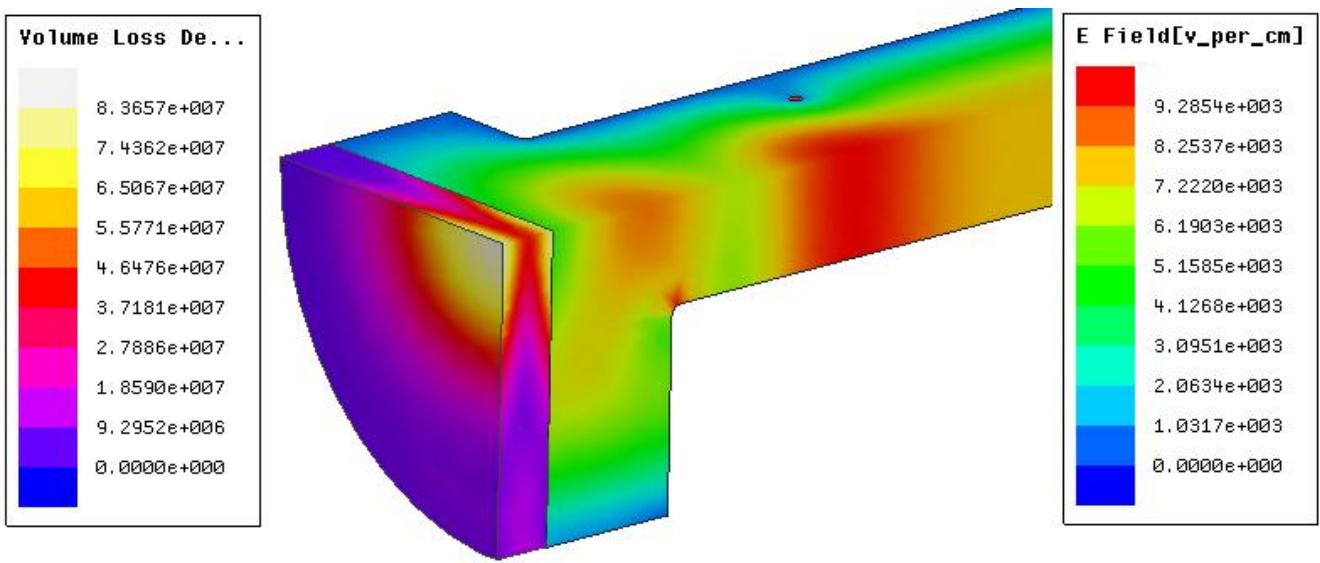

**Figure 2. Volume loss density in the ¼ ceramic (left scale in W/m$^3$) and electric field norm (right scale in V/cm) for 500 kW input on matched load.**

In order to conform to ITER acceptance criteria for vacuum feedthroughs [24], the maximum pressure on the ceramic (air) has been specified to be 3 bars absolute (0.3 MPa). The water cooling circuit specifications has been specified in accordance with the NFRI 5 GHz klystron test-bed in which the high power tests have been performed. The maximum pressure on the cooling circuit has been set to 6 bars, with a maximum pressure drop on the cooling circuit for a 8 L/min flow of 0.3 bar. It has to be noted that these specifications are well below the ITER port plug specifications, for which the water temperature and pressure are expected to be 70-100°C up to 40 bars in cooling mode and up to 250°C and 45 bars in baking mode However, such conditions are challenging because the copper skirt thickness has to be increased to sustain such a pressure. Increasing the skirt thickness in order to withstand a 40-45 bars pressure will reduce the cooling efficiency. Moreover, if the skirt is thicker, the mechanical constraints on the ceramic due to the differential thermal expansion between the BeO and Cu will be more severe since the skirt will be stiffer. For this reason, the present proposed prototype design only focuses on the RF performances.

Assuming an inlet flow of 8 L/min and an outlet return pressure of 1.5 bar specified by the water cooling loop system, the fluid velocity calculated from computational fluid dynamics (CFD) modelling in Ansys Fluent at the 10 mm diameter inlet is 1.7 m/s. Figure 3 illustrates the water velocity in half of the cooling section. The flow in the fluid section is mainly turbulent, as the Reynold number at the inlet/outlet is of the order of 10000 and 40000 in the skirt. Since inlet and outlet have been located normal to the skirt, the water velocity rise up to 4.6 m/s at the interface between the pipe and the skirt. However, the flow is rapidly homogenous in the skirt and with lower velocities (cf. Figure 4), leading to an homogenous cooling of the ceramic. The average value of the heat transfer coefficient is $H=12300$ $W.m^{-2}K^{-1}$ on the inner side of the skirt. The water pressure is homogenous inside the skirt as illustrated in Figure 5 and the estimated pressure drop is 0.12 bar. Cavitation regime is not expected in the cooling section, as the cavitation number is more than two orders of magnitude above the critical threshold (Ca ~ 1) in all situations.

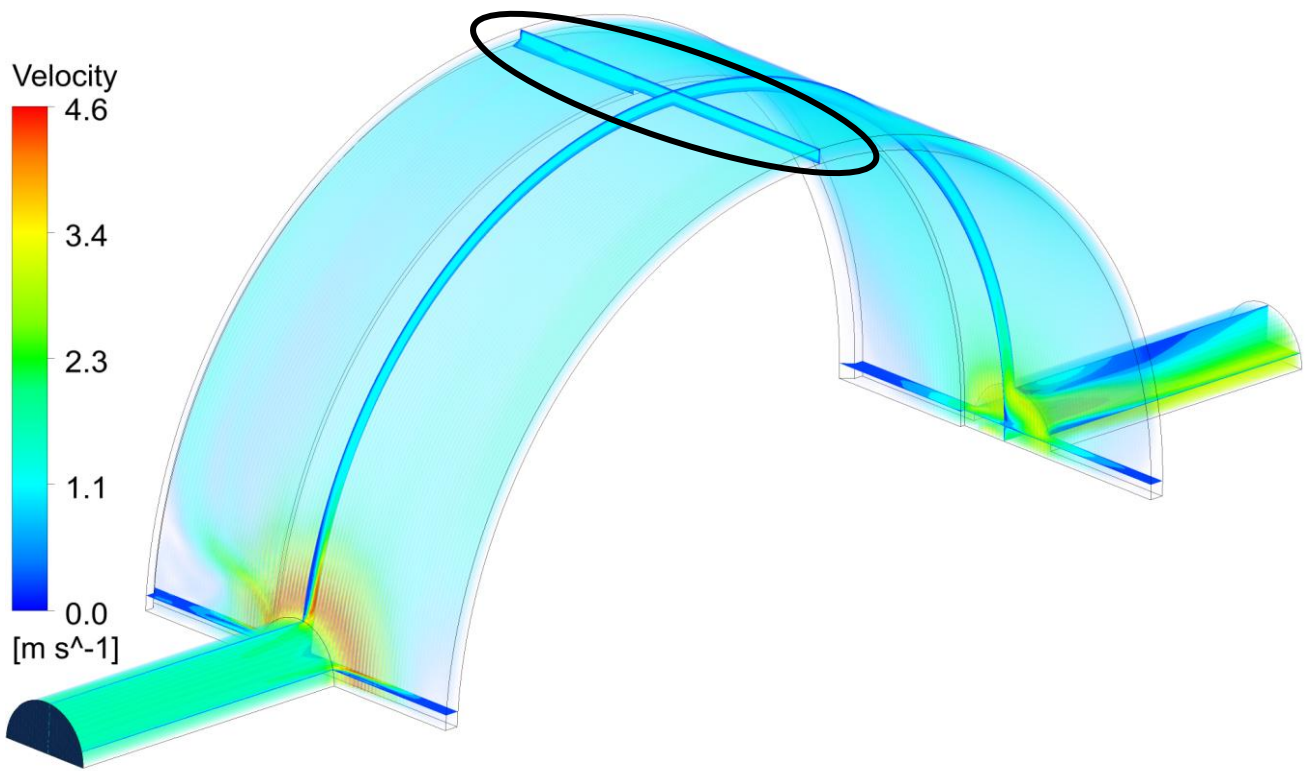

**Figure 3. Water velocity in the skirt around the ceramic. Inlet is located at the left of the picture. The black oval illustrates the close-up view of Figure 4.**

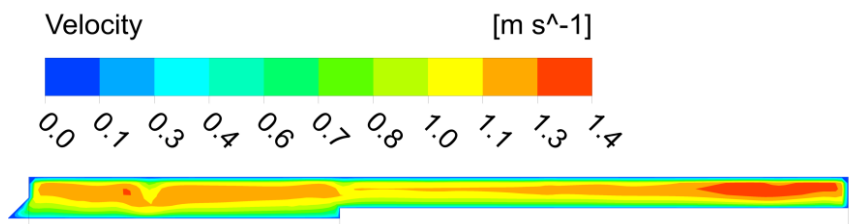

**Figure 4. Close-up of the fluid velocity in the skirt section.**

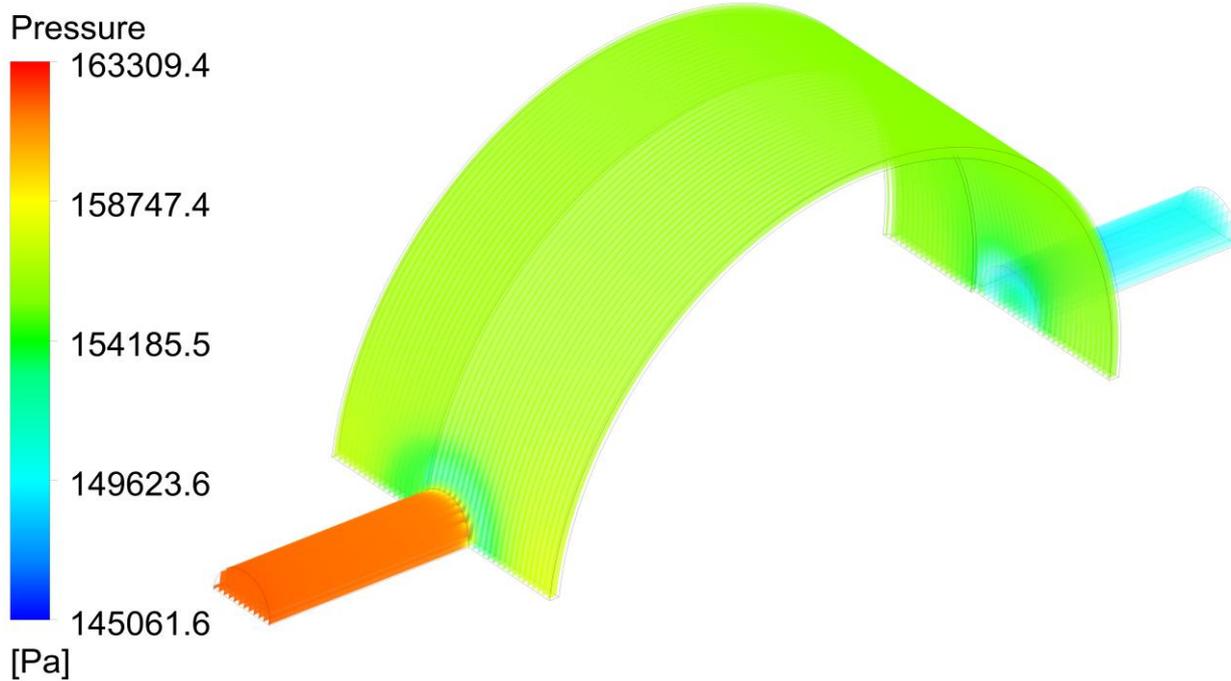

**Figure 5. Water pressure in the skirt around the ceramic. Inlet is located at the left of the picture.**

The thermal model of the window has been set-up using loading from both the electromagnetic (heat load on walls and ceramic) and CFD models (convection on the skirt). Illustrations of the resistive losses and the dielectric losses for 500 kW input are given in Figure 6 and Figure 7.

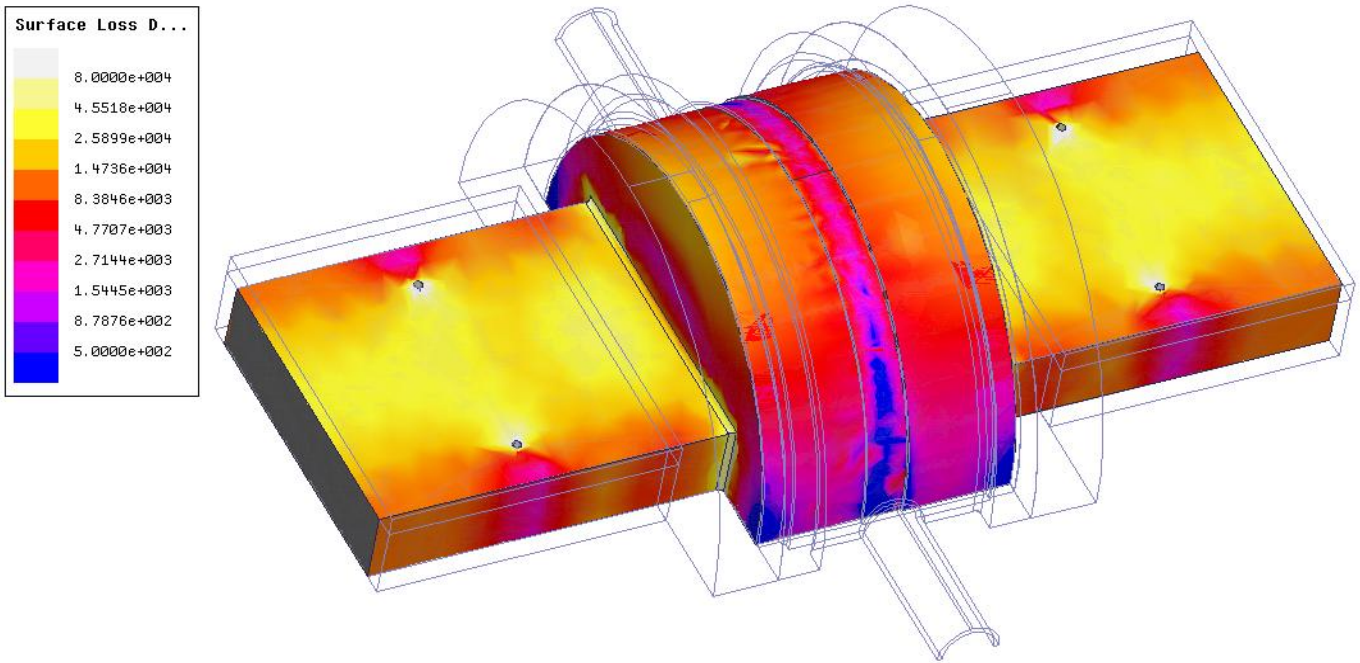

**Figure 6. RF Resistive Heat Flux for 500 kW input power in W/m². The surface integrated loss on the conductor is of the order of 720 W with a pessimistic copper conductivity of σ=44 MS/m.**

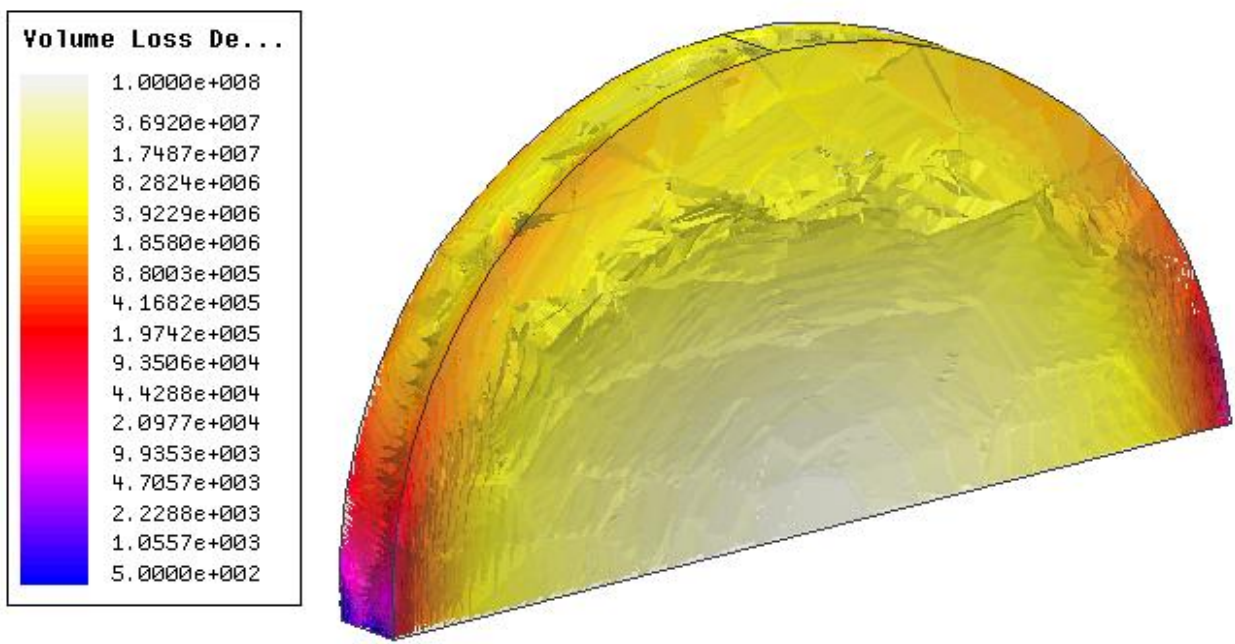

**Figure 7. Dielectric losses distribution in W/m³ in the ceramic for 500 kW input on matched load. The volume integrated loss in the ceramic is of the order of 550 W.**

A first thermal steady-state study has been conducted to survey the maximum reached temperature. The thermal loads come from the RF losses in the ceramic and on the waveguides walls. The thermal sinks come from water cooling and air radiation. The rectangular waveguide parts and the skirt are modelled as OFHC copper. The rectangular to circular waveguide transitions are in Stainless-steel. Conduction losses on copper or copper-coated waveguide walls are estimated from Ansys/HFSS using a pessimistic conductivity of σ=44 MS/m. The dielectric losses in the ceramic are also calculated from full-wave modelling and coupled to the thermal model. The connections between the elements, especially between the BeO and the copper skirt, are modelled as perfect (bonded). The mechanical parameters of the BeO used in the thermal model are reported in Table 1. In the results below, the thermal conductivity of the BeO has been kept constant with respect to the temperature. Some studies indicate however that the thermal conductivity drops with temperature [27], which would increase the BeO temperature.

| | |
|---:|:---|
| Density | 2850 kg/m$^{-3}$ |
| Thermal Expansion | 8e-6 K$^{-1}$ |
| Specific Heat | 250 J/(kg.K) |
| Thermal Conductivity | 265 W/(m.K) |

**Table 1. BeO 99.5% Mechanical and Thermal Properties (Source: American Beryllia Product Specification Sheet EPS-8001).**

As the matching rods are not directly cooled by the water circuit, their temperatures reach 180°C after a 5 seconds pulse and nearly 290°C in steady-state regime. For this reason, the nominal RF pulse duration of these prototypes has been limited to be 5 seconds as a conservative approach. For a CW design, cooling pipes on the upper and lower parts of the rectangular waveguides would solve the problem.

In the ceramic, the maximum temperature is reached at the centre where the dielectric loss density is the maximum and which is also the farthest point from the water heat sink. Starting from an initial water temperature of 22°C, the ceramic centre temperature is calculated to be 75°C after a 5 s/500 kW pulse. The initial temperature of the ceramic is recovered after nearly 10 s, as illustrated in Figure 8. In order to limit the mechanical stresses on the ceramic, the temperature difference between the centre and the periphery of the disk must be kept under a safety limit of 80°C. This limit has been determined by both mechanical modelling and rule of thumb, in order to keep stresses in the ceramic below the static fatigue limit [8].

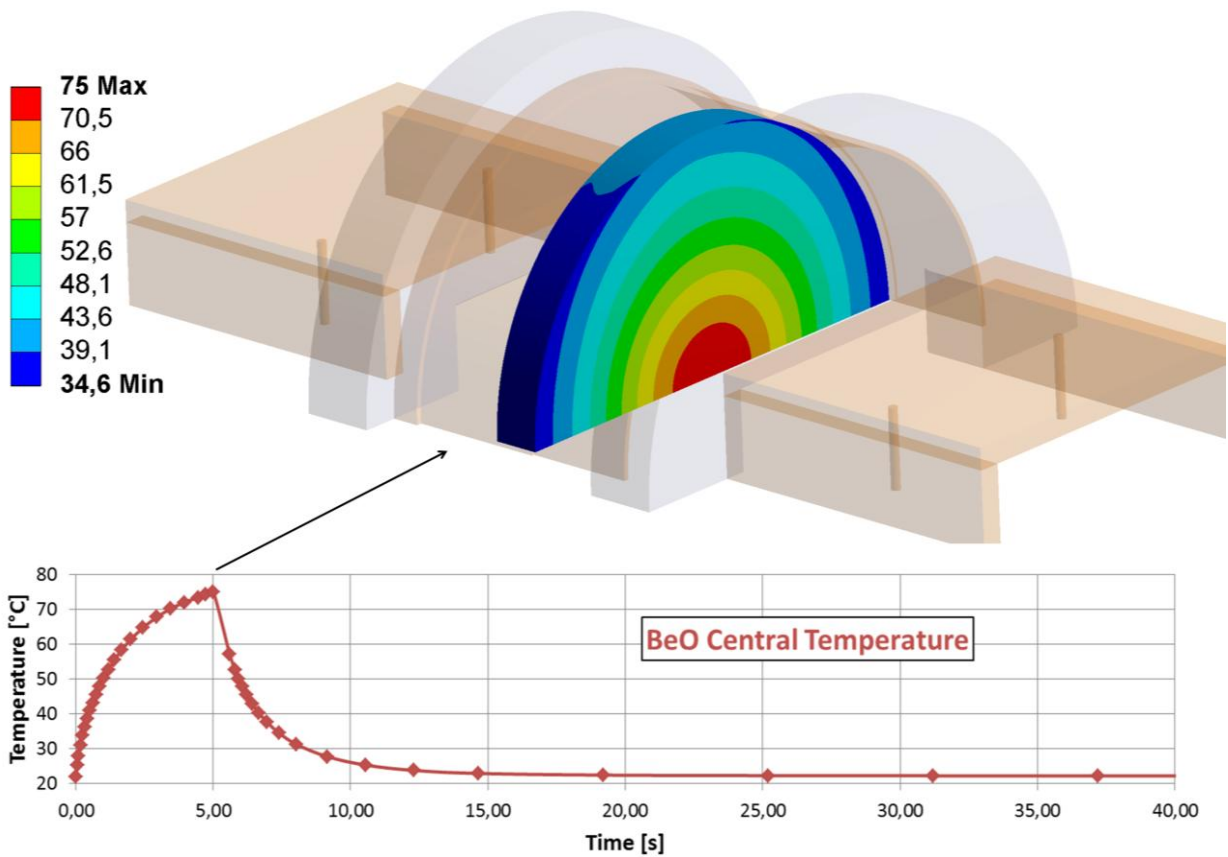

**Figure 8. Modelling of the temperature evolution of the BeO ceramic during a 5 s/ 500kW RF pulse.**

## 3. Manufacturing and Low Power Tests

Two RF windows have been manufactured by the PMB/ALCEN Company in collaboration with CEA/IRFM. For ensuring the possibility to perform ITER relevant tests, like high temperature baking up to 240°C [24] , the window flanges are non-standard and have been designed to be compatible for both Viton© or Helicoflex© seals. As shown in **FIGURE 9** (b), a rectangular groove is machined on both sides of the window. Male waveguide adapters are then required to connect the window: these adapters are designed to insure a good electrical contact at the flange by a heel slightly longer than the groove depth, in order to dissociate the electrical contact from the sealing aspects. In return, additional elements must be connected together, which can lead to additional losses in the transmission line.

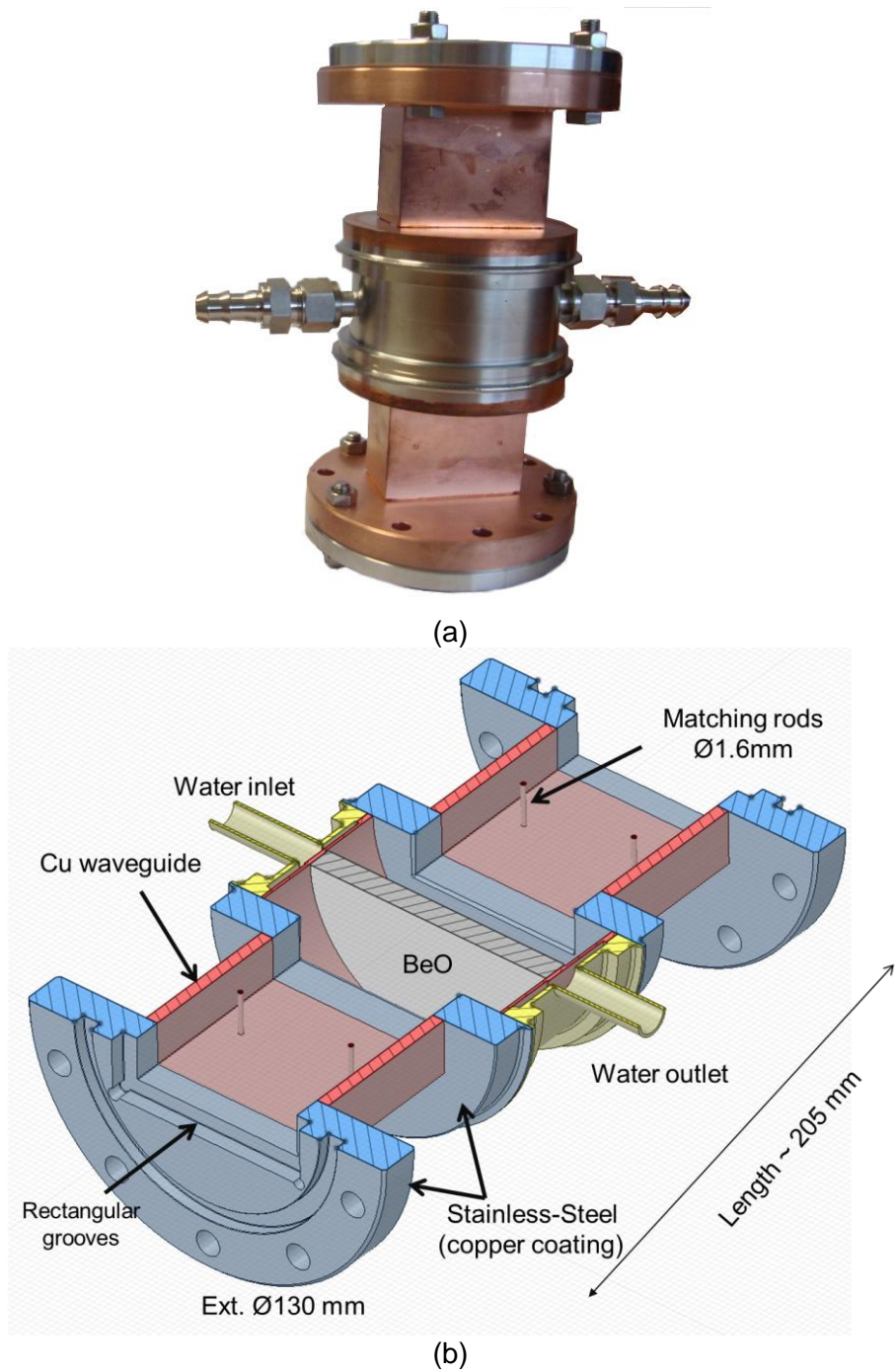

**FIGURE 9.** (a): picture of one window. (b): CAD cut-view of the window.

BeO ceramic has a high secondary electron emission coefficient, which enhances multipactor induced breakdown probability. In order to suppress the multipactor induced breakdown at the surface of the ceramic, a 10 nm titanium nitride (TiN) coating is applied by vapour deposition. This coating intends to decrease the secondary emission yield of the ceramic and to improve the electrical charge exchange [22] . Stainless-steel parts facing the RF fields are copper coated (10 to 20 µm). After brazing the rectangular copper waveguides elements together, the stainless-steel parts are TiG welded.

Low power RF measurements (dBm level) are reported in **FIGURE 11** for both windows, labelled #1 and #2. The return losses ($S_{11}$ and $S_{22}$) at 5 GHz are below -32 dB (VSWR < 1.05:1) in accordance with the RF modelling. The optimum frequency is slightly shifted to low frequencies by 20 to 40 MHz, for reasons that will be explained in the last section of this paper. It has to be noted that the return losses $S_{11}$ and $S_{22}$ are not symmetrical for both windows, differently from the results of the RF modelling. This fact is attributed to both the measurement errors (which come from RF cables) and the matching rods which were deformed during manufacturing. These deformations, due to a handling problem during the copper coating operation, have been experienced for both windows. Even if the rods have been reshaped with specific tooling as best as possible, slight mechanical differences exist between each rods (at the contrary of the RF model which is ideal and perfectly symmetric).

The insertion losses ($S_{21}$) are reported in **FIGURE 11**. They have been measured at 5 GHz in the range from -0.01 dB to -0.05 dB (i.e. between 0.23% and 1.14% of input power). The overall uncertainty of the transmission losses measurements is of the order of 0.03 dB. These measurements reflected not only the window transmission losses but also the part of the assembly for which no calibration was possible (such as the dedicated waveguides adapters with specific flanges). Precise measurements of the insertion losses are very sensitive to the test bed mechanical assembly, in particular to the quality of the mechanical assembly of all flanges.

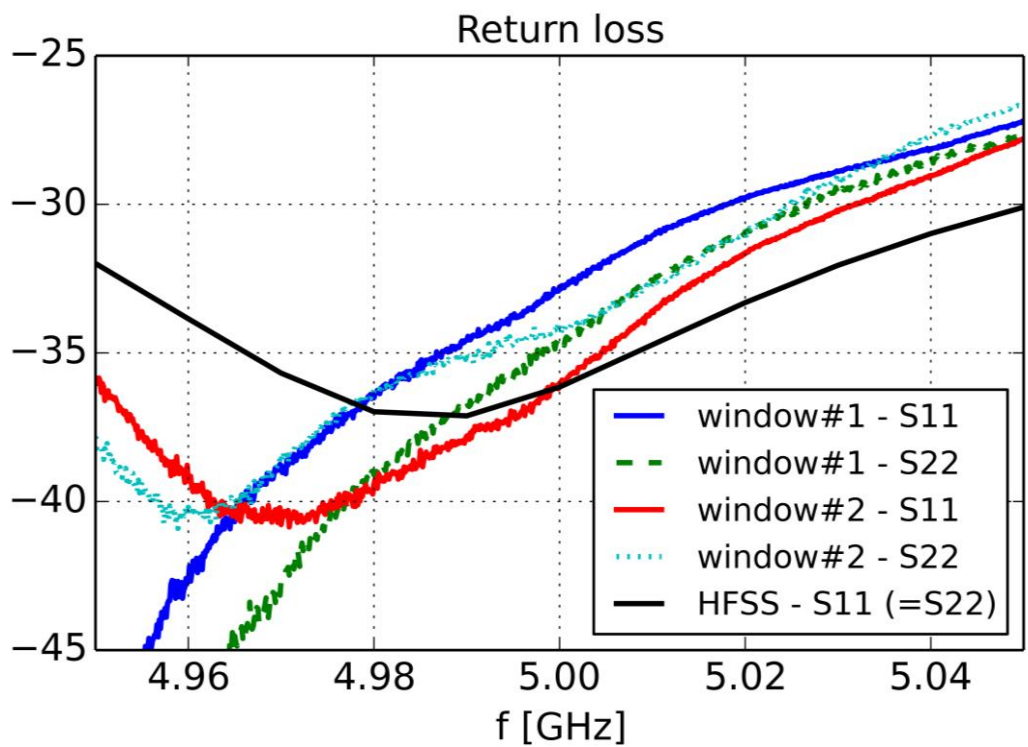

**FIGURE 10**. Return loss [dB] for both windows. RF modelling prediction(plain black) from HFSS.

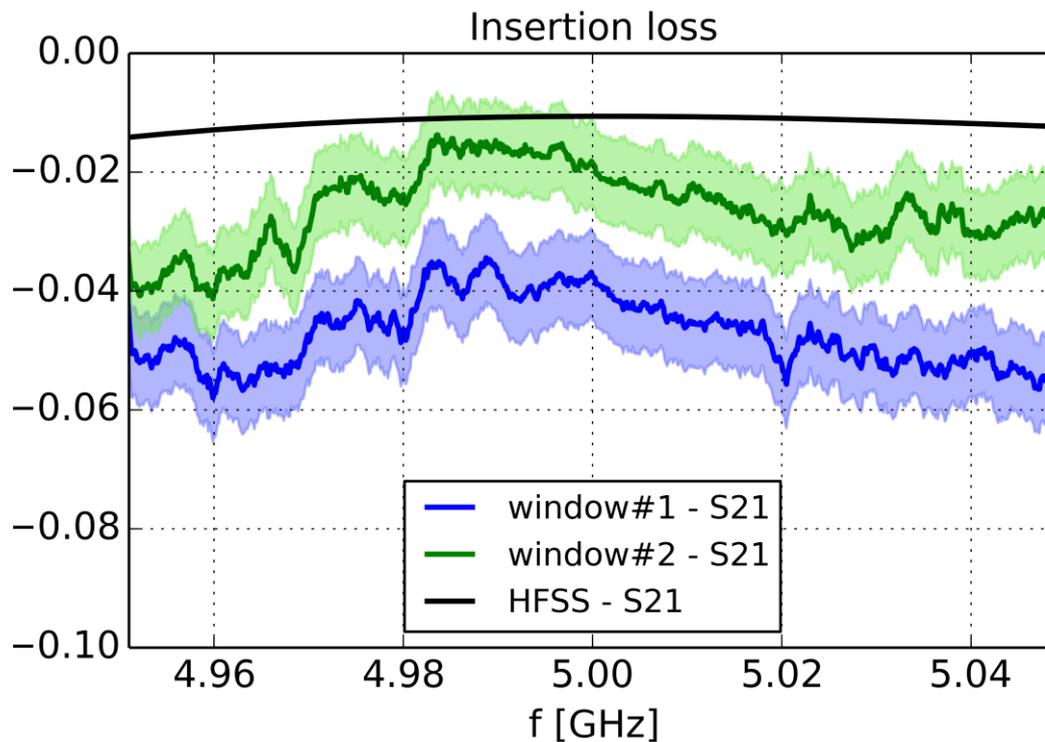

**FIGURE 11**. Insertion loss [dB] for both windows. RF modelling prediction(plain black) from HFSS. Bold line represent the averaged values and shaded areas the minimum and maximum of the measurement noise.

## 4. High power RF tests

High power measurements have been performed on the NFRI LH test bed in 2013 and 2014. The setup shown in **FIGURE 12** consists in a single window inserted in the transmission line. The transmission line is filled with $SF_6$ gas. A vacuum test with both windows at the same time using an interspace pumping system was also expected in case of successful completion of the tests of both prototypes. The windows were cooled with 8 to 17 l/min water flow at 20-25°C. Calorimetric measurements of the window water cooling loop were recorded during shots and used to estimate the power losses in the window [15]. The temperature of the ceramic core was monitored by an infra-red camera (Agema thermovision 900, spectral range 2-5.4 µm) during 500 kW RF power shots. The measured BeO temperature depends on the BeO emissivity which is unknown. An absolute temperature calibration has been performed between 24.3°C to 81.7°C using electrical blanket heating and shows a linear trend. Higher temperatures are deduced by a linear interpolation of the calibration curve.

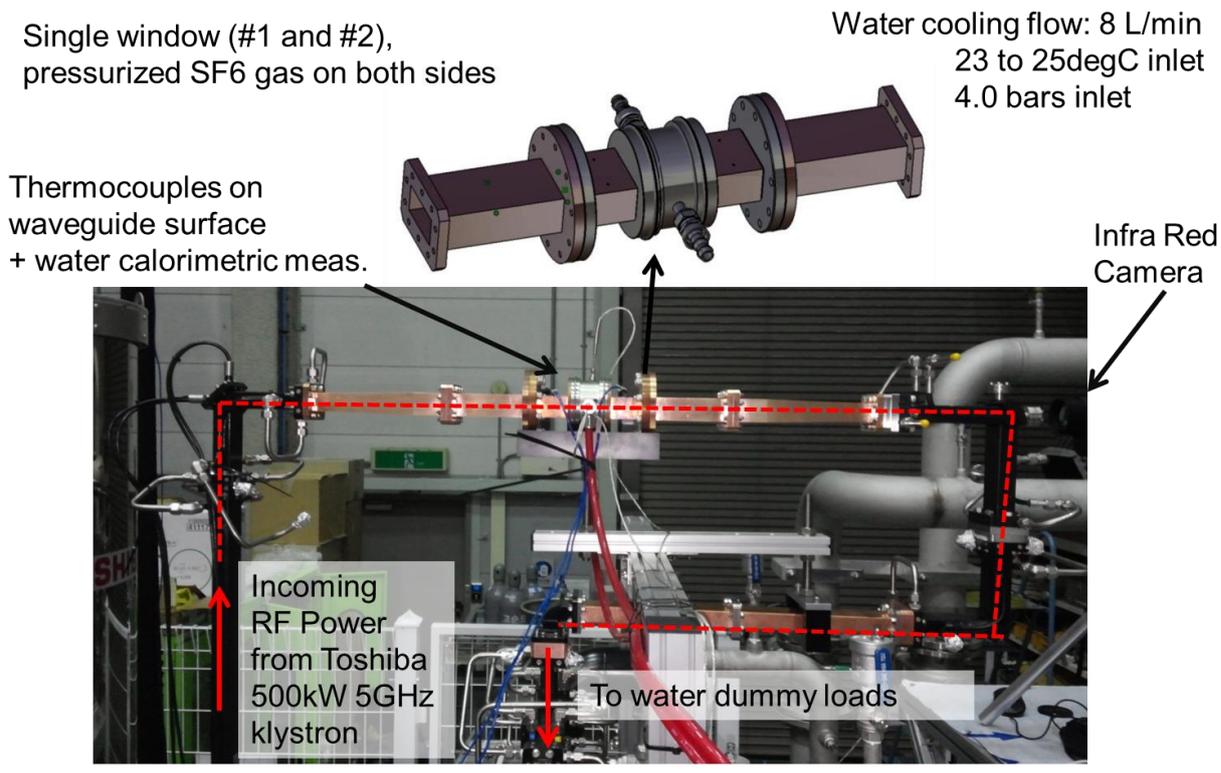
**FIGURE 12.** High power test setup at 5 GHz/500kW NFRI test bed in Korea.

The RF energy applied for both windows is illustrated in Figure 13 as a function of the pulse number. The RF power has been applied during single pulses length of 100 ms starting from 25 kW, and then the power was increased progressively pulse by pulse by 25 kW steps up to 250 kW. RF pulse length has been increased progressively (in more than 30 pulses) up to 2.5 s at a power of 250 kW. The RF power has been increased up to 500 kW during 100 ms pulses, then the pulse length have been increased to few seconds at 500 kW, in more than 50 pulses. Repetitive pulses at 500 kW have been performed for different pulse lengths. A single breakdown event has been detected via the optical fiber protection system at 250 kW. The fraction of reflected power measured close to the klystron output ranged from -22.6 dB to -21.6 dB (VSWR 1.16:1 to 1.18:1) for the first window and from -21.2 dB to -20.8 dB (VSWR 1.19:1 to 1.2:1) for the second window. It has to be noted that the return losses are inclusive of the return losses produces by both the window and the whole assembly, including its various components such as bends and water loads, for which a -24.9 dB (VSWR 1.12:1) return loss has been measured without the window installed in the transmission line.

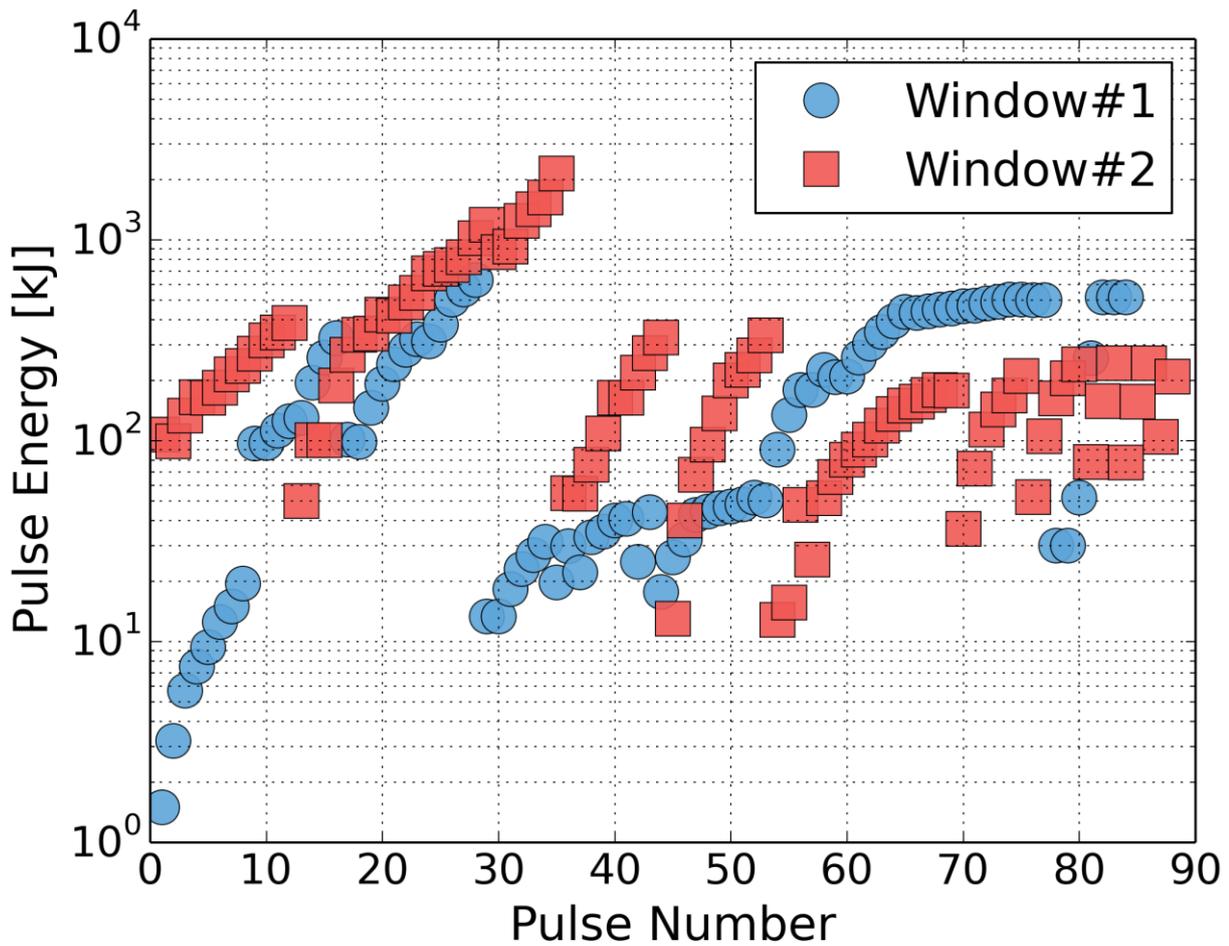

**Figure 13. Pulse Energy in kJ as a function of the pulse number for both windows.**

The water temperature increase ΔT during the RF power application from calorimetric measurement indicated a power loss in the device from 0.89% to 1.1% of the input power for the first window and of 0.7% for the second window. This power loss includes the ceramic but also all the neighbouring copper-plated waveguide sections at both sides of the window, which were not water-cooled during these tests. This value is well above the 0.2% expected from RF modelling or best measured values, but is consistent with the lowest range of values from low-power measurements reported in section 2. These values are also in the same range of the ones measured on the Toshiba 5 GHz RF windows which equipped the klystron[26] . Total loss versus input power for both windows is reported in Figure 14 and shows linear trends. Two trend curves are related to the first window (1.1% and 0.89% of the incident power $P_{in}$) and the third one to the second window (0.73% of $P_{in}$). The total losses are larger in the first window than in the second window, in consistence with low power measurements. The difference in the trends for the first window is not explained and is thought to come from measurement dispersion. These linear trends exclude eventual non-linear loss processes during these short-length pulses (up to 5 s).

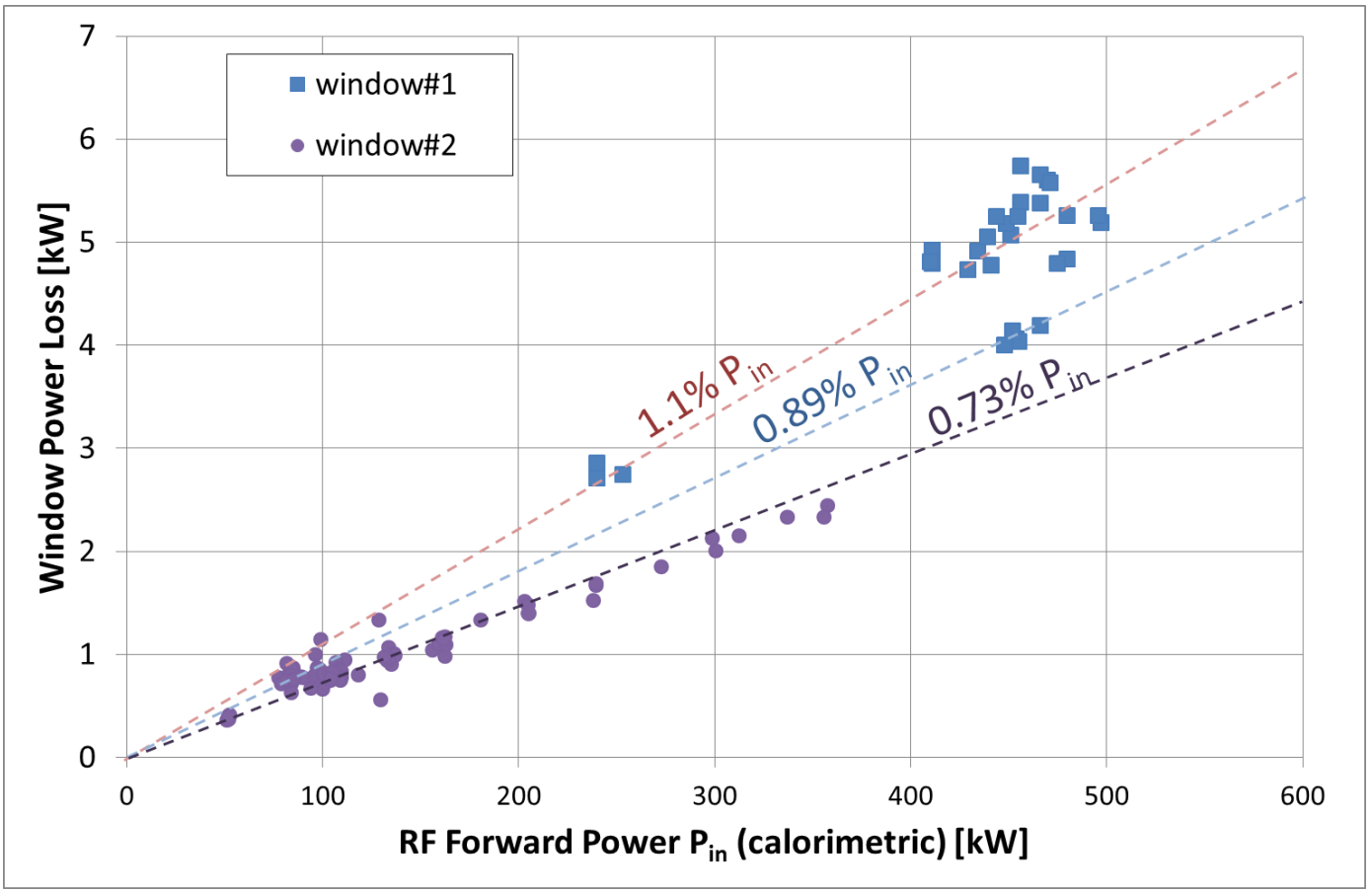

**Figure 14.** Measured calorimetric power losses for both prototype windows VS forward RF power Pin from dummy loads calorimetric measurement during the different experimental days.

Infra-red (IR) measurements of the ceramic temperature have been made during high power pulses. The rods small diameter (1.6 mm) makes its temperature measurement difficult because of the camera resolution, but also mainly because of the IR reflections occurring on them. When the ceramic temperature is increasing, the rods become undistinguishable from the ceramic. Some thermocouples have been placed at the top of the window (cf **FIGURE 12**), located specifically over the matching rods, but no change of temperature during the pulses was measured, in agreement with modelling. The ceramic central temperature (maximum temperature), measured by the IR camera after 3 seconds pulses at 500 kW (Figure 15), is recurrently higher than 250 °C, and thus higher than the thermal modelling prediction illustrated in Figure 8 which was below 100°C after a 5 s pulse.

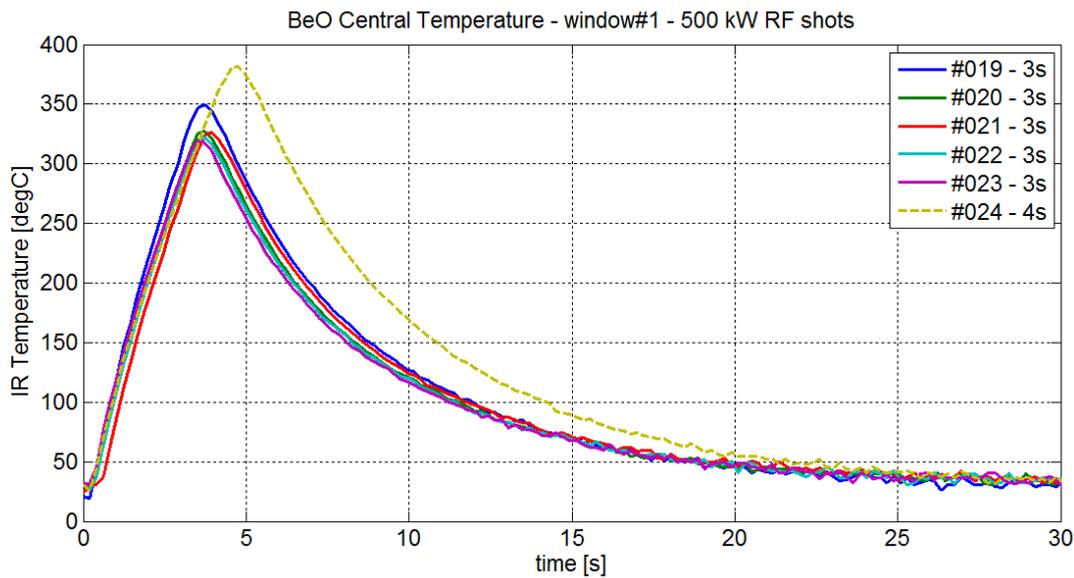

Figure 15. BeO Central temperature evolution 3 s and 4 s shots, measured by the IR Camera for 500 kW±20kW input shots for window #1.

Moreover, the cooling time constant is around twice larger than expected from ideal thermal modelling, indicating that a thermal barrier reduces the heat conduction efficiency. The losses are not affected by a degradation of the cooling efficiency, since the content of lost power does not depend on the thermal conductivity if we assume that the dielectric loss tangent does not change with the temperature increase.

The fact that the BeO central temperature increases to temperatures higher than 300°C prevent these two windows prototype to be used in CW regime at full power without exceeding the mechanical stress limits of the ceramic. The first ceramic window prototype was found to be cracked after disassembly from the test-bed. Following this observation, the maximum power and the pulse length has been reduced for the following tests of the second window, in order not to exceed a maximum IR measured temperature of 100°C to avoid high temperature gradient between the ceramic core and its periphery cooled by the 20-25°C water loop.

## 5. Experimental Results Analysis

Following the high power tests, two 16 mm diameter samples have been machined from the first (cracked) BeO ceramic window. A post-mortem RF analysis of these two samples has been performed with the same methodology used for the manufacturer sample (Cf. Section 2). The relative permittivity measured on these two pieces is 6.74±0.12 and 6.72±0.01, to be compared with the 6.36 value originally expected from the manufacturer sample measurement. The tan δ was also up to twice higher than expected from the BeO sample measurement (from $5.4 \times 10^{-4}$ up to $8.8 \times 10^{-4} \pm 1.1 \times 10^{-5}$). RF modelling shows that the change of relative permittivity has little effect on the dielectric losses, indicating that no other higher order modes are excited in the ceramic. The change of permittivity shifts the optimum frequency by -20 MHz, which corroborates low power measurements (**FIGURE 10**). A loss tangent value of $9\times10^{-4}$ leads to double the total losses in the ceramic. Doubling the dielectric loss tangent into an ideal thermal model leads to a BeO central temperature of nearly 140°C after a 5 s pulse. However, even taking into account these changes in the reference RF model, the predicted total RF losses reach 2 kW, still far from the 5 kW and 3.5 kW measured by calorimetry for window #1 and #2 respectively.

We observe in Figure 14 that the total losses are almost linear with respect to the input power, thus excluding any significant changes on the dielectric loss tangent or copper conductivity with temperature for these short-length pulses. As the power lost in the ceramic is linearly linked to the dielectric loss tangent, the latter is thought to be the dominant player. The effect of the reflected power from the waveguide bends and the water loads has been assessed and is marginal in the total RF losses evaluation. Adding small gaps between the bonded parts of the window or at the flange interfaces may possibly causes additional RF losses. However, a reliable estimate of these losses is difficult.

As seen in **FIGURE 12**, the window is connected to copper waveguide adapters on both sides. Each side is the addition of two waveguide sections of 140 mm and 248 mm length respectively. As these waveguides attached to the window were not actively cooled, a fraction of the RF power lost into them participates to the power balance measured by the calorimetric method. With a realistic copper conductivity of $\sigma=44$ MS/m, the conduction losses for WR229 waveguides at 5 GHz for 500 kW are 2.97 kW/m. Taking into account only the two 140 mm length waveguides, the power balance leads to an additional RF power loss of 830 W. Taking into account all the waveguides length (140 mm + 248 mm) the power balance leads to an additional RF power loss of 2300 W. However, estimating the real fraction of power lost in these waveguides and dissipated into the window water cooling loop is delicate because of the large uncertainty of the thermal contacts between flanges.

Taking into account the post-mortem BeO RF properties in the RF model leads to an insertion loss of -0.02 dB (0.5%) and -0.03 dB (0.7%) if one includes the connection waveguides on each sides of the window. Modelling and measurements are then consistent if one assumes that between 5 and 20 % of the power measured by calorimetry is dissipated in the connection waveguides. These modelling however do not take into account the fact that the matching rods are slightly deformed.

The longer characteristic time measured from IR during cooling can be originated from either a decreased BeO thermal conductivity, a reduced water heat transfer coefficient or a thermal barrier between the BeO and the copper skirt. If the first two have little effect on the thermal modelling results, unless using non-realistic values, the latter is thought to be the main cause of a reduced cooling efficiency. The brazing layer is made of gold/copper alloys (Au50/Cu50) in order to avoid silver alloys, which are undesirable in ITER because of transmutation risks during the nuclear phase. This brazing material has a low thermal conductivity (in the order of 33 W/m/K). Moreover, in order to increase the wettability of the BeO to the brazing alloy, a metallization of the BeO periphery is performed using a Molybdenum/Manganese layer, with the addition of nickel plating. Both have also a low thermal conductivity. Taking into account the Molybdenum/Manganese layer which has an electrical conductivity about 6 MS/m in the RF model as an impedance boundary condition leads to increase the RF losses around the ceramic of about 10 W for 500 kW input. Thermal modelling of such brazing layers, which thickness is typically between 20 to 40 µm, is delicate and is highly sensitive to parameter values. However, the general trend when applying a degraded thermal contact at the interface between the BeO and the copper is a fast increase of the temperature rise time due to a thermal barrier effect and similarly an increase of the characteristic cooling time, both observed experimentally.

## 6. Conclusion and perspectives

In the frame of a R&D effort conducted by CEA toward the design and the qualification of a 5 GHz LHCD system for the ITER tokamak, two 5 GHz 500 kW/5 s windows have been designed and tested at high power in collaboration with NFRI. The window design relies on a symmetrical pill-box concept with a cylindrical beryllium oxide (BeO) ceramic brazed on an actively water cooled copper skirt. The ceramic RF properties have been measured on a manufacturer test sample in order to get realistic values for guiding the design. Low power measurements show return losses below -32 dB (VSWR < 1.05:1) and insertion losses between -0.01 dB and -0.05 dB (i.e. 0.23% to 1.14% of input power), with the optimum frequency shifted toward frequencies lower than the nominal one. High power tests conducted at NFRI show a total power loss of 1.1% of the input power for one window and of 0.7% for the second one, in agreement with the lowest range of cold RF measurement but not with initial RF simulations. These values are also in the same range than the ones measured on the two Toshiba 5 GHz RF windows which equipped the klystron. The ceramic temperature during RF pulses has been found to reach unexpected high values, preventing these windows to be used under CW conditions because of the ceramic stress limits.

A post-mortem RF analysis of two samples of one of the windows shows that the dielectric relative permittivity of the ceramic was 6.7 while for the sample initially measured it was 6.36. In addition, the measured dielectric loss tangent on the two post-mortem ceramic samples is up to twice as high as the manufacturer sample. The difference in the permittivity explains the shift of the optimum frequency. Additional RF modelling including the new dielectric properties of BeO, a more realistic RF conduction loss rate on copper and the additional connection waveguides, shows that the calculated RF losses match the observations. Additional losses can also be thought to come from flanges misalignments and small gaps. However, the only increase of the dielectric loss is not sufficient to explain the high temperatures measured. The higher temperature and delay time is explained by a thermal barrier due to ceramic metallization and the use of low thermal conduction brazing materials.

This shows that large margins must be taken from the reference design (or alternatively accept large production wastes). This aspect is especially challenging in the case of 500 kW 5 GHz windows, since the power density is large and leads to high temperature and mechanical stresses, close to acceptable margins. For future developments, avoiding matching rods would simplify the mechanical design, but may also suppress the possibility to make any post-manufacturing tuning on the window. Keeping a temperature gradient between the ceramic centre and its periphery as low as possible in addition insure the lowest losses as possible are the main design goal and would prevent from overstress in the ceramic. Other mechanical design improvements such as for the water pipes inlet/outlet geometry could lead to decrease pressure drop and increase the water velocity in the skirt and thus improve the heat transfer efficiency with the ceramic.

## 7. Acknowledgements

The authors would like to thank Olivier Tantot and Damien Passerieux from the XLIM Laboratory in Limoges, France, where the dielectric measurements on BeO samples have been conducted.